\documentclass[twocolumn,showpacs,preprintnumbers,superscriptaddress,prb,floatfix,aps,10pt]{revtex4-1}

\usepackage{amsmath,amssymb}
\usepackage{graphicx,textcomp}
\usepackage{subfigure}

\newcommand{\figref}[1]{Fig.~\ref{#1}}
\renewcommand {\vec}    [1]    {\ensuremath{\mathbf{#1}}}

\newcommand   {\mat}    [1]    {\ensuremath{\mathbf{\bar{\bar{#1}}}} }			
\newcommand   {\pderiv} [2]    {\ensuremath{\frac{\partial#1}{\partial#2}}}		

\begin{document}

\title{Temperature dependent effective third order interatomic force constants from first principles}
\author{Olle Hellman}
\affiliation{Department of Physics, Chemistry and Biology (IFM), Link\"oping University, SE-581 83, Link\"oping, Sweden.}

\author{I. A. Abrikosov}
\affiliation{Department of Physics, Chemistry and Biology (IFM), Link\"oping University, SE-581 83, Link\"oping, Sweden.}

\begin{abstract}
The temperature dependent effective potential (TDEP) method is generalized beyond pair interactions. The second and third order force constants are determined consistently from ab initio molecular dynamics simulations at finite temperature. The reliability of the approach is demonstrated by calculations of the Mode Gr\"uneisen parameters for Si. We show that the extension of TDEP to higher order allows for an efficient calculation of the phonon life time, in Si as well as in $\epsilon$-FeSi, a system that exhibits anomalous softening with temperature.
\end{abstract}

\maketitle

\section{Introduction}

Thermodynamic properties of materials are often discussed in terms of the quasiharmonic approximation.\cite{Born1998,Maradudin1968} This theory has a solid foundation, but it is not without limitations. Any property that relies on phonon lifetimes and scattering rates, such as thermal conductivity, is unavailable. For those properties one needs terms higher than second order in the Taylor expansion of the crystal potential energy surface. If the higher order terms are known, there is extensive theory developed for the properties that can be extracted.\cite{srivastava1990physics} The difficulties lie in determining these parameters.

Density functional theory gives one access to the potential energy surface. Perturbation theory and the $2n+1$ theorem, or direct supercell approaches can be used to determine materials force constants.\cite{Broido2007,Lindsay2012,Narasimhan1991} In these formalisms, however, the potential energy surface is treated as constant with respect to temperature. We have previously shown that this is not the case.\cite{Hellman2011,Hellman2013} In our previous, using the temperature dependent effective potential (TDEP), we obtain the best possible second order Hamiltonian as a fit to the Born-Oppenheimer molecular dynamics potential energy surface at finite temperature. With this technique it is possible, for example, to accurately describe solid $^4$He, which is strongly anharmonic, with a second order Hamiltonian. The effective potential gives accurate phonon dispersion relations and free energies. The aim of this paper is to extend the TDEP formalism to include higher order terms, making the technique suitable for calculations of important materials properties such as phonon lifetimes.

\section{The method}

We start with a model crystal Hamiltonian:
\begin{equation}\label{eq:harmhamiltonian}
\begin{split}
H= & U_0+\sum_i \frac{m\vec{p}_i^2}{2}+
\frac{1}{2!}\sum_{ij\alpha\beta}\Phi_{ij}^{\alpha\beta}
u_i^\alpha u_j^\beta \\
& +\frac{1}{3!}
\sum_{ijk\alpha\beta\gamma}\Psi_{ijk}^{\alpha\beta\gamma}
u_i^\alpha u_j^\beta u_k^\gamma.
\end{split}
\end{equation}
Here $U$ is the potential energy, $\mat{\Phi}_{ij}$ and $\mat{\Psi}_{ijk}$ are the second and third order force constants. The displacement of atom $i$ from ideal positions is denoted $\vec{u}_i$, momentum of at and $\alpha\beta\gamma$ are Cartesian indices. Bold symbols indicate vectors and doubly overlined symbols matrices or tensors respectively.

The basic idea of the generalized TDEP is to use Born-Oppeheimer molecular dynamics to accurately sample the potential energy surface at finite temperature. Then we fit the model Hamiltonian in Eq. \ref{eq:harmhamiltonian} to this surface. This is done by comparing the forces of the model and the \emph{ab initio} system at each time step and minimizing the difference.

With the vast number of values to be determined for the third order force constants it is crucialfor the generalization of the TDEP to higher order terms to make use of the symmetry analysis.\cite{Hellman2013} We begin by reiterating the symmetry relations the force constants obey,\cite{Maradudin1968} first the transposition symmetries:
\begin{align}
\label{eq:firstsym}\Phi_{ij}^{\alpha\beta} &= \Phi_{ji}^{\beta\alpha} \\
\Psi_{ijk}^{\alpha\beta\gamma} = \Psi_{ikj}^{\alpha\gamma\beta} = \Psi_{jik}^{\beta\alpha\gamma}& =
\Psi_{jki}^{\beta\gamma\alpha} =
\Psi_{kij}^{\gamma\alpha\beta} =
\Psi_{kji}^{\gamma\beta\alpha}.
\end{align}
Then, if two tensors are related by symmetry operation $S$ the components are related as follows:
\begin{align}
\Phi_{ij}^{\alpha\beta} &= 
\sum_{\mu\nu}\Phi_{kl}^{\mu\nu}
S^{\mu\alpha}S^{\nu\beta}  \\
\Psi_{ijk}^{\alpha\beta\gamma} &= 
\sum_{\mu\nu\xi}\Psi_{mno}^{\mu\nu\xi}
S^{\mu\alpha} S^{\nu\beta} S^{\xi\gamma}.
\end{align}
Force constants also obey the acoustic sum rules:
\begin{align}
\sum_j \mat{\Phi}_{ij} & =0 \quad \forall\, i \\
\label{eq:lastsym}\sum_k \mat{\Psi}_{ijk} & =0 \quad \forall\, i,j
\end{align}
To apply symmetry relations \eqref{eq:firstsym}-\eqref{eq:lastsym}, we set each tensor component to a symbolic variable, called $\theta_k$. The index $k$ runs from 1 to the total number of components in all tensors. We include all tensors within a cutoff radius $r_c$ (the maximum cutoff is determined by the simulation cell size). Using Eqs. \ref{eq:firstsym}--\ref{eq:lastsym} we figure out which tensor components that are equal, related to each other or 0 by symmetry. This drastically reduces the number of values that have to be determined. With the symmetry irreducible representation at hand, we express the forces in the model Hamiltonian:
\begin{equation}\label{eq:forceeq0}
F_i^\alpha=\sum_{j\beta}\Phi_{ij}^{\alpha\beta}
u_j^\beta
+\frac{1}{2}
\sum_{jk\beta\gamma}\Psi_{ijk}^{\alpha\beta\gamma}
u_j^\beta u_k^\gamma.
\end{equation}
Evaluating \eqref{eq:forceeq0} analytically allows of to express the forces as a function of the symmetry inequivalent components $\theta_k$:
\begin{equation}\label{eq:forceeq1}
F_i^\alpha=\sum_k \theta_k c_k^{i\alpha}(\vec{U}).
\end{equation}
Here $c_k^{i\alpha}(\vec{U})$, the coefficient for each $\theta_k$ is a polynomial function of all displacements within $r_c$. The form of this function depends on the crystal at hand. For a given supercell, we can express Eq. \ref{eq:forceeq1} as a matrix equation:
\begin{equation}
\vec{F}_M=\vec{\Theta}\mat{C}(\vec{U}).
\end{equation}
The subscript $M$ denotes the forces from the model potential. The coefficent matrix $\mat{C}$ is a function of all the displacements in the supercell. $\vec{\Theta}$ is a vector holding all the $\theta_k$. To obtain a solution for $\vec{\Theta}$ we run Born-Oppenheimer molecular dynamics in the canonical ensemble at temperature $T$. From these simulations, we store displacements $\vec{u}$ and forces $\vec{F}_{MD}$ at each time step. Then, we seek the $\vec{\Phi}$ that minimize the difference between the model system and the \emph{ab initio} one:
\begin{equation}\label{eq:min_f_phi}
\begin{split}
\min_{\Theta}\Delta \vec{F} & =\frac{1}{N_t} \sum_{t=1}^{N_t}  \left| \vec{F}_t^{\textrm{MD}}-\vec{F}_t^{\textrm{H}} \right|^2= \\
& =\frac{1}{N_t} \sum_{t=1}^{N_t} \left| \vec{F}_t^{\textrm{MD}}-\mat{C}(\vec{U}_t^{\textrm{MD}})\vec{\Theta} \right|^2 = \\
& = \frac{1}{N_t} \left\Vert 
\begin{pmatrix} \vec{F}_1^{\textrm{MD}} \\  \vdots \\ \vec{F}_{N_t}^{\textrm{MD}} \end{pmatrix}-
\begin{pmatrix} \mat{C}(\vec{U}_1^{\textrm{MD}}) \\ \vdots \\ \mat{C}(\vec{U}_{N_t}^{\textrm{MD}}) \end{pmatrix}\vec{\Theta}
\right\Vert^2.
\end{split}
\end{equation}
Here $N_t$ is the number of time steps in the molecular dynamics, and subscript $t$ denotes the displacements and forces from time step $t$. A least square solution,
\begin{equation}\label{eq:phi_eqFC}
\vec{\Theta}=
\begin{pmatrix}
\mat{C}(\vec{U}_1^{\textrm{MD}}) \\
\vdots \\
\mat{C}(\vec{U}_{N_t}^{\textrm{MD}})
\end{pmatrix}^{+}
\begin{pmatrix}
\vec{F}^{\textrm{MD}}_1 \\
\vdots \\
\vec{F}^{\textrm{MD}}_{N_t}
\end{pmatrix}.
\end{equation}
gives the $\vec{\Theta}$ that minimizes these forces. Then, with a simple substitution back into $\Phi_{ij}$ and $\Psi_{ijk}$ we determine the quadratic and cubic force constants. Note that the second and third order force constants are extracted from the same set of displacements and forces, simultaneously. 

\section{Numerical results}

\begin{figure}
\includegraphics[width=\linewidth]{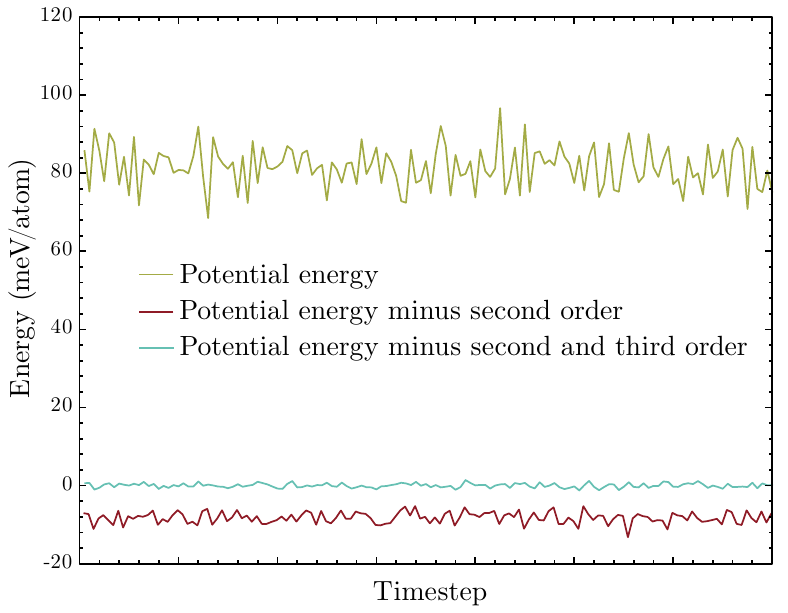}
\caption{\label{fig:epotdiff} (color online) Figure showing the difference between the \emph{ab initio} potential energies of Si and the potential energies from our model potential. The top line are the values from Born-Oppenheimer molecular dynamics. The bottom line is the \emph{ab initio} energies with the second order term in Eq. \ref{eq:harmhamiltonian} subtracted, and the middle line is when both the second and third order energies are subtracted. Had the model potential been exact, the middle line would be perfectly straight. There are still some fluctuations, but they are on the order of 1meV/atom, and the accuracy of our potential is good. The second order potential includes up to 4th neighbours and the third order up to 2nd nearest neighbours.}
\end{figure}

\begin{figure}
\includegraphics[width=\linewidth]{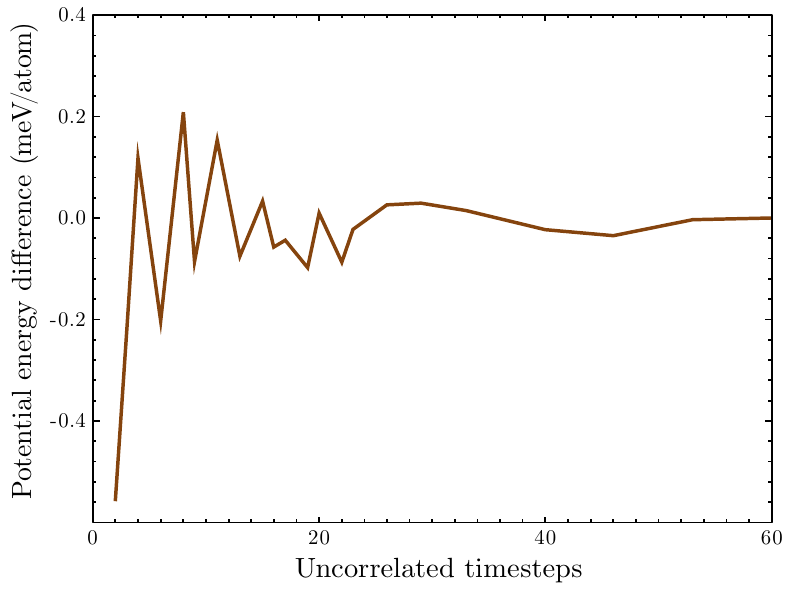}
\caption{\label{fig:epotconv}(color online) Convergence of the potential energy of our model potential for Si. The displacements that go into Eq. \ref{eq:harmhamiltonian} are from snapshots from molecular dynamics.}
\end{figure}

Si is a common model system where high order terms are relevant,\cite{Omini,Pascual-Gutierrez2009,Broido2007,Narasimhan1991} and to verify the quality of the force constants obtained in the TDEP formalism we employ Born-Oppenheimer molecular dynamics with the projector-augmented wave (PAW) method as implemented in the code VASP.\cite{Kresse1999,Kresse1996,Kresse1993b,Kresse1996c}
We use a 128 atom supercell. For the BZ integration we use the $\Gamma$-point and ran the simulations on a grid of temperatures and volumes in the NVT ensemble. Temperature was controlled using a Nos\'e thermostat.\cite{Nose1984} Exchange-correlation effects were treated using the generalized gradient approximation with the Perdew--Burke--Ernzerhof\cite{Perdew1996} functional. We use a plane wave cutoff of 250 eV. The simulations ran for about 100ps with a time step of 1 fs. A subset of uncorrelated samples is then chosen. For each of the samples the electronic structure and total energies are recalculated using a $5\times5\times5$ k-point grid and a cutoff of 500eV.

The first way of verifying if our force constants are correct, is to calculate the potential energy according to Eq. \ref{eq:harmhamiltonian} and compare to the ones from DFT. In \figref{fig:epotdiff} we show the difference in potential energy from DFT and the model Hamiltonian, and in \figref{fig:epotconv} we show the convergence of this potential energy with respect to the number of timesteps. These results confirm that the third order force constants are accurately determined and represent the potential energy surface well.

\begin{figure}
\includegraphics[width=\linewidth]{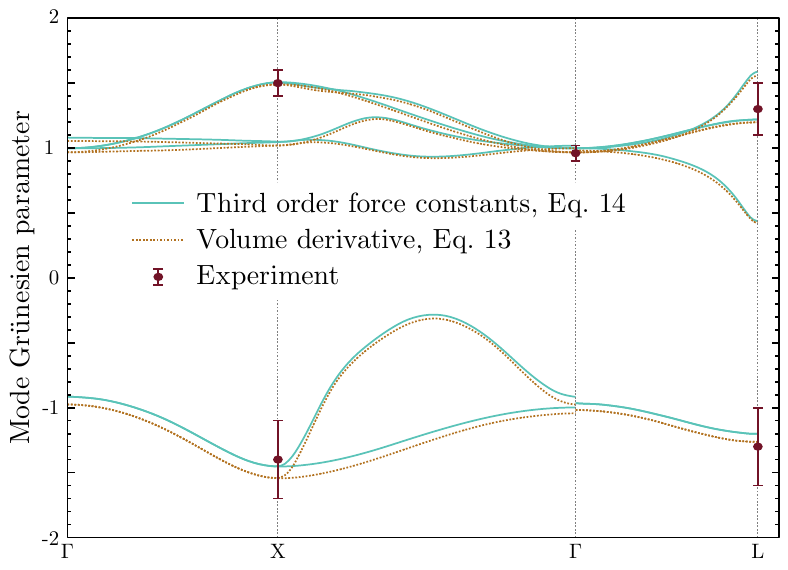}
\caption{\label{fig:sigrun}(color online) Mode Gr\"uneisen parameters for Si. The solid line is calculated according to Eq. \ref{eq:grun2} and the dotted line according to Eq. \ref{eq:grun1}. The experimental points are from Weinstein \emph{et al.}\cite{Weinstein1975}.}
\end{figure}

The mode Gr\"uneisen parameters are a measure how sensitive the vibrational frequencies are with respect to a volume change. They are given by
\begin{equation}\label{eq:grun1}
\gamma_{\vec{q}s}=-\frac{V}{\omega_{\vec{q}s}}\pderiv{\omega_{\vec{q}s}}{V}
\end{equation}
where $V$ is the volume and $\omega_{\vec{q}s}$ is the frequency of mode $s$ at wave vector $\vec{q}$. $\gamma_{\vec{q}s}$ can be obtained either by numerical differentiation of the phonon dispersion relations or from the third order force constants\cite{Fabian1997,Broido2005}:
\begin{equation}\label{eq:grun2}
\gamma_{\vec{q}s}=-\frac{1}{6\omega_{\vec{q}s}^2}\sum_{ijk\alpha\beta\gamma}
\frac{\epsilon_{i\alpha}^{\vec{q}s*} \epsilon_{j\beta}^{\vec{q}s}}
{\sqrt{M_i M_j}}
r_k^\gamma \Psi_{ijk}^{\alpha\beta\gamma}e^{i\vec{q}\cdot\vec{r}_j}
\end{equation}
Here $\epsilon_{i\alpha}^{\vec{q}s}$ is component $\alpha$ associated eigenvector $\epsilon$ for atom $i$. $M_i$ is the mass of atom $i$, and $\vec{r}_i$ is the vector locating its position. To confirm the accuracy and consistency of our third order force constants, we calculated the mode Gr\"uneisen parameters using both Eq. \ref{eq:grun1} and \ref{eq:grun2}, as can be seen in \figref{fig:sigrun}, the results are excellent, both in terms of consistency with each other and with respect to the experimental values. 

\begin{figure}
\includegraphics[width=\linewidth]{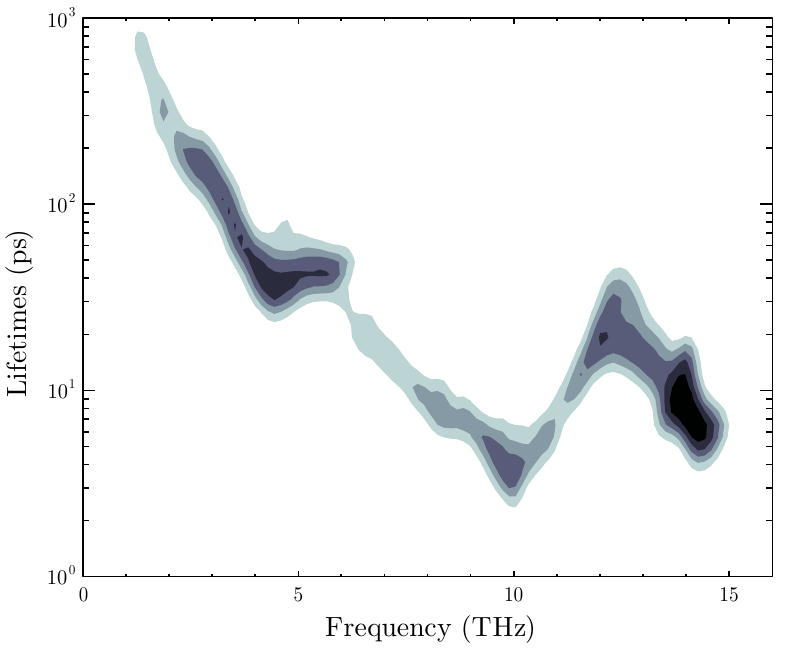}
\caption{\label{fig:silifetimes}(color online) Density plot of phonon lifetimes in Si at 300K. The intensity is logarithmic.}
\end{figure}

With the third order force constants we can calculate the phonon lifetimes in the relaxation time approximation. The lifetime due to phonon-phonon scattering is related to the imaginary part of the phonon self energy:\cite{wallace1998thermodynamics}
\begin{equation}
\frac{1}{\tau_{\vec{q}s}}=\Gamma_{\vec{q}s}
\end{equation}
where $\tau_{\vec{q}s}$ is the lifetime for wave vector $\vec{q}$ and mode $s$, and
\begin{equation}
\begin{split}
\Gamma_{\vec{q}s}=& \sum_{s's''} \frac{\hbar \pi}{16} \iint_{\mathrm{BZ}}
\left|\Psi^{\vec{q}\vec{q}'\vec{q}''}_{ss's''}\right|^2 \Delta_{\vec{q}\vec{q}'\vec{q}''} \times \\
& \bigl[(n_{\vec{q}'s'}+n_{\vec{q}''s''}+1)
\delta(\omega_{\vec{q}s}-\omega_{\vec{q}'s'}-\omega_{\vec{q}''s''}) \\
+ & 2(n_{\vec{q}'s'}-n_{\vec{q}''s''})
\delta(\omega_{\vec{q}s}-\omega_{\vec{q}'s'}+\omega_{\vec{q}''s''}) \bigr]d\vec{q}'d\vec{q}''.
\end{split}
\end{equation}
$n_{\vec{q}s}$ is the equilibrium occupation number. The $\Delta_{\vec{q}\vec{q}'\vec{q}''}$ ensures momentum conversation, $\vec{q}+\vec{q}'+\vec{q}''=\vec{G}$, and the deltafunctions in frequency ensure energy conservation. The three-phonon matrix elements are given by
\begin{equation}
\begin{split}
\Psi^{\vec{q}\vec{q}'\vec{q}''}_{ss's''} = &
\sum_{ijk}
\sum_{\alpha\beta\gamma}
\frac{
\epsilon^{\vec{q}s}_{\alpha i} \epsilon^{\vec{q}'s'}_{\beta j} \epsilon^{\vec{q}''s''}_{\gamma k}
}{
\sqrt{M_{i}M_{j}M_{j}}
\sqrt{\omega_{\vec{q}s}\omega_{\vec{q}'s'}\omega_{\vec{q}''s''}}
}\times \\ &
\Psi^{\alpha\beta\gamma}_{ijk}
e^{i \vec{q}\cdot\vec{r}_1 + i \vec{q}'\cdot\vec{r}_2+i \vec{q}''\cdot\vec{r}_3}
\end{split}
\end{equation}
where $M_i$ is the mass of atom $i$, $\epsilon^{\vec{q}s}_{\alpha i}$ is component $\alpha$ of the eigenvector for mode $\vec{q}s$ and atom $i$.
The resulting phonon lifetimes for Si can be seen in \figref{fig:silifetimes}. The numerical integration is done on a $31 \times 31 \times 31$ Monkhorst-Pack\cite{Monkhorst1976a} q-point grid. The momentum conservation is exactly fulfilled (the sum of two vectors on the grid ends up on the grid), and for the energy conservation we used the adaptive broadening scheme of \citet{Yates2007}. Our results agree well with previously calculated results.\cite{Pascual-Gutierrez2009,Esfarjani2011}

\section{Temperature dependence}

\begin{figure}
\includegraphics[width=\linewidth]{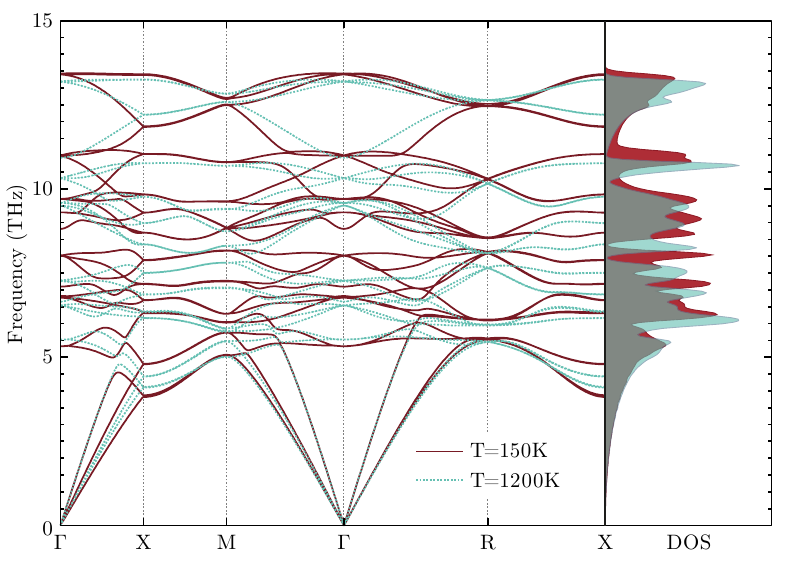}
\caption{\label{fig:fesidisprel}(color online) FeSi dispersion relations and phonon density of states. The solid lines correspond to T=150K and the dashed lines to T=1200K. The experimentally observed softening across the spectrum can be seen clearly.}
\end{figure}

The unique feature of our formalism if that the force constants are volume and temperature dependent. For comparison, only the volume dependence is included in the quasiharmonic approximation. In systems with dynamical instabilities, such as bcc Zr, the temperature dependence is obvious,\cite{Hellman2011} but it is also present in system that do not exhibit instabilities. $\epsilon$-FeSi is such a system.\cite{Delaire2011a} 

\begin{figure}[t!]
\includegraphics[width=\linewidth]{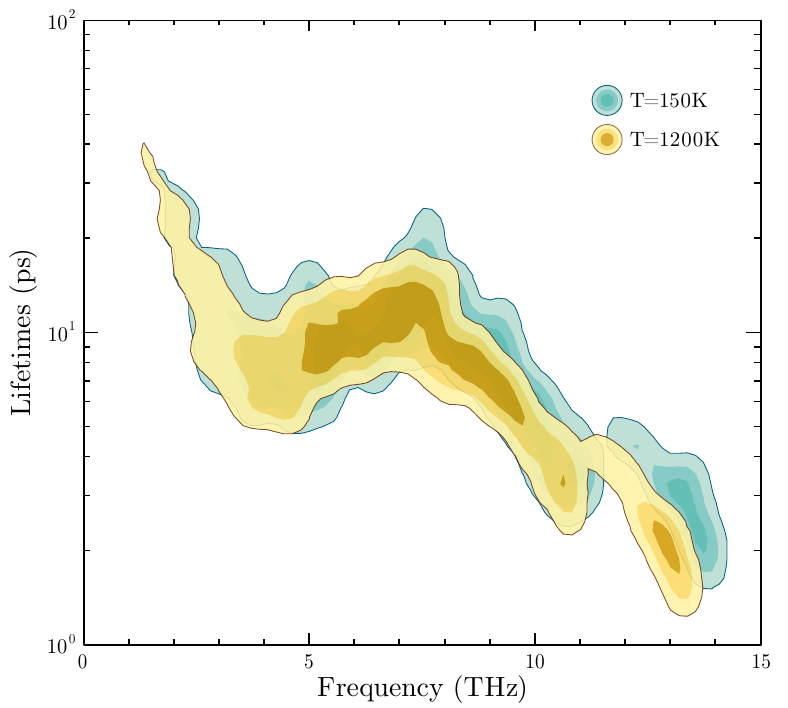}
\caption{\label{fig:fesigrun}(color online) FeSi phonon lifetimes. The liftetimes are evaluated at 400K using force constants extracted from 150K and 1200K.}
\end{figure}

We ran Born-Oppenheimer molecular dynamics for FeSi out in a similar fashion to that of Si, but we use a $3\times3\times3$ supercell (216 atoms), a cutoff of 300eV and used the $\Gamma$-point for Brillouin zone integration. We used the experimental lattice parameter of 4.779Å and temperatures of 150 and 1200K and ran the simulations for 30ps after equilibration with a 1fs time step. We used the same lattice constant at both temperatures to show that the softening is not due to thermal expansion, but originates from finite temperature effects, such as electron-phonon coupling.

Phonon dispersion relations from TDEP calculations are shown in \figref{fig:fesidisprel}. We observe the softening across the whole spectrum with increasing temperature. The physical origin is discussed by \citet{Delaire2011a} in great detail. It exhibits anomalous softening with temperature due to the thermal exitations smearing a sharp peak at the Fermi level, inducing a insulator-metal transition. We observe that TDEP can capture this temperature dependence well. The third order force constants are also temperature dependent, the result of this can be seen in \figref{fig:fesigrun}. The lifetimes are decreased significantly, across the whole spectrum. The temperature parameter used when evaluating lifetimes was fixed at 400K, the only difference comes from the temperature dependence of the force constants. This is consistent with experimental results,\cite{Nyhus1995} where they see a strong suppression of phonon linewidths below 250K. Due to the overestimation of the band gap in DFT, the closing of the gap occurs at a higher temperature in simulations (it is not fully closed until 1200K\cite{Delaire2011a}), but the effect on the phonon linewidths and dispersions is qualitatively correct.

This tells us that to accurately describe the phonon-phonon interactions of FeSi at finite temperature, one needs to take the temperature dependence of the potential energy surface into account.

\section{Conclusions}

We have presented an extension of our existing formalism to calculate temperature dependent third order force constants. They are shown to reproduce experimental results well. This is a numerically efficient technique to simultaneously incorporate all orders of phonon-phonon and electron phonon coupling into a model Hamiltonian.

\section{Acknowledgements}

Support from the Knut \& Alice Wallenberg Foundation (KAW) project ``Isotopic Control for Ultimate Material Properties'', the Swedish Research Council (VR) projects 621-2011-4426 and LiLi-NFM, the Swedish Foundation for Strategic Research (SSF) program SRL10-0026, and the Ministry of Education and Science of the Russian Federation within the framework of Program ``Research and Pedagogical Personnel for Innovative Russia (2009-2013)'' (project no. 14.В37.21.0890 of 10.09.2012) is gratefully acknowledged. SSI acknowledges support from the Swedish Government Strategic Research Area Grant in Materials Science to AFM research environment at LiU. Supercomputer resources were provided by the Swedish National Infrastructure for Computing (SNIC).



\begin{thebibliography}{25}%
\makeatletter
\providecommand \@ifxundefined [1]{%
 \@ifx{#1\undefined}
}%
\providecommand \@ifnum [1]{%
 \ifnum #1\expandafter \@firstoftwo
 \else \expandafter \@secondoftwo
 \fi
}%
\providecommand \@ifx [1]{%
 \ifx #1\expandafter \@firstoftwo
 \else \expandafter \@secondoftwo
 \fi
}%
\providecommand \natexlab [1]{#1}%
\providecommand \enquote  [1]{``#1''}%
\providecommand \bibnamefont  [1]{#1}%
\providecommand \bibfnamefont [1]{#1}%
\providecommand \citenamefont [1]{#1}%
\providecommand \href@noop [0]{\@secondoftwo}%
\providecommand \href [0]{\begingroup \@sanitize@url \@href}%
\providecommand \@href[1]{\@@startlink{#1}\@@href}%
\providecommand \@@href[1]{\endgroup#1\@@endlink}%
\providecommand \@sanitize@url [0]{\catcode `\\12\catcode `\$12\catcode
  `\&12\catcode `\#12\catcode `\^12\catcode `\_12\catcode `\%12\relax}%
\providecommand \@@startlink[1]{}%
\providecommand \@@endlink[0]{}%
\providecommand \url  [0]{\begingroup\@sanitize@url \@url }%
\providecommand \@url [1]{\endgroup\@href {#1}{\urlprefix }}%
\providecommand \urlprefix  [0]{URL }%
\providecommand \Eprint [0]{\href }%
\providecommand \doibase [0]{http://dx.doi.org/}%
\providecommand \selectlanguage [0]{\@gobble}%
\providecommand \bibinfo  [0]{\@secondoftwo}%
\providecommand \bibfield  [0]{\@secondoftwo}%
\providecommand \translation [1]{[#1]}%
\providecommand \BibitemOpen [0]{}%
\providecommand \bibitemStop [0]{}%
\providecommand \bibitemNoStop [0]{.\EOS\space}%
\providecommand \EOS [0]{\spacefactor3000\relax}%
\providecommand \BibitemShut  [1]{\csname bibitem#1\endcsname}%
\let\auto@bib@innerbib\@empty
\bibitem [{\citenamefont {Born}\ and\ \citenamefont {Huang}(1964)}]{Born1998}%
  \BibitemOpen
  \bibfield  {author} {\bibinfo {author} {\bibfnamefont {M.}~\bibnamefont
  {Born}}\ and\ \bibinfo {author} {\bibfnamefont {K.}~\bibnamefont {Huang}},\
  }\href@noop {} {\emph {\bibinfo {title} {{Dynamical theory of crystal
  lattices}}}}\ (\bibinfo  {publisher} {Oxford University Press},\ \bibinfo
  {address} {Oxford},\ \bibinfo {year} {1964})\BibitemShut {NoStop}%
\bibitem [{\citenamefont {Maradudin}\ and\ \citenamefont
  {Vosko}(1968)}]{Maradudin1968}%
  \BibitemOpen
  \bibfield  {author} {\bibinfo {author} {\bibfnamefont {A.~A.}\ \bibnamefont
  {Maradudin}}\ and\ \bibinfo {author} {\bibfnamefont {S.}~\bibnamefont
  {Vosko}},\ }\href {\doibase 10.1103/RevModPhys.40.1} {\bibfield  {journal}
  {\bibinfo  {journal} {Reviews of Modern Physics}\ }\textbf {\bibinfo {volume}
  {40}},\ \bibinfo {pages} {1} (\bibinfo {year} {1968})}\BibitemShut {NoStop}%
\bibitem [{\citenamefont {Srivastava}(1990)}]{srivastava1990physics}%
  \BibitemOpen
  \bibfield  {author} {\bibinfo {author} {\bibfnamefont {G.~P.}\ \bibnamefont
  {Srivastava}},\ }\href {http://books.google.se/books?id=OE-bHd2gzVgC} {\emph
  {\bibinfo {title} {{The Physics of Phonons}}}}\ (\bibinfo  {publisher} {A.
  Hilger},\ \bibinfo {year} {1990})\BibitemShut {NoStop}%
\bibitem [{\citenamefont {Broido}\ \emph {et~al.}(2007)\citenamefont {Broido},
  \citenamefont {Malorny}, \citenamefont {Birner}, \citenamefont {Mingo},\ and\
  \citenamefont {Stewart}}]{Broido2007}%
  \BibitemOpen
  \bibfield  {author} {\bibinfo {author} {\bibfnamefont {D.~a.}\ \bibnamefont
  {Broido}}, \bibinfo {author} {\bibfnamefont {M.}~\bibnamefont {Malorny}},
  \bibinfo {author} {\bibfnamefont {G.}~\bibnamefont {Birner}}, \bibinfo
  {author} {\bibfnamefont {N.}~\bibnamefont {Mingo}}, \ and\ \bibinfo {author}
  {\bibfnamefont {D.~a.}\ \bibnamefont {Stewart}},\ }\href {\doibase
  10.1063/1.2822891} {\bibfield  {journal} {\bibinfo  {journal} {Applied
  Physics Letters}\ }\textbf {\bibinfo {volume} {91}},\ \bibinfo {pages}
  {231922} (\bibinfo {year} {2007})}\BibitemShut {NoStop}%
\bibitem [{\citenamefont {Lindsay}\ \emph {et~al.}(2012)\citenamefont
  {Lindsay}, \citenamefont {Broido},\ and\ \citenamefont
  {Reinecke}}]{Lindsay2012}%
  \BibitemOpen
  \bibfield  {author} {\bibinfo {author} {\bibfnamefont {L.}~\bibnamefont
  {Lindsay}}, \bibinfo {author} {\bibfnamefont {D.}~\bibnamefont {Broido}}, \
  and\ \bibinfo {author} {\bibfnamefont {T.}~\bibnamefont {Reinecke}},\ }\href
  {\doibase 10.1103/PhysRevLett.109.095901} {\bibfield  {journal} {\bibinfo
  {journal} {Physical Review Letters}\ }\textbf {\bibinfo {volume} {109}},\
  \bibinfo {pages} {1} (\bibinfo {year} {2012})}\BibitemShut {NoStop}%
\bibitem [{\citenamefont {Narasimhan}\ and\ \citenamefont
  {Vanderbilt}(1991)}]{Narasimhan1991}%
  \BibitemOpen
  \bibfield  {author} {\bibinfo {author} {\bibfnamefont {S.}~\bibnamefont
  {Narasimhan}}\ and\ \bibinfo {author} {\bibfnamefont {D.}~\bibnamefont
  {Vanderbilt}},\ }\href {\doibase 10.1103/PhysRevB.43.4541} {\bibfield
  {journal} {\bibinfo  {journal} {Physical Review B}\ }\textbf {\bibinfo
  {volume} {43}},\ \bibinfo {pages} {4541} (\bibinfo {year}
  {1991})}\BibitemShut {NoStop}%
\bibitem [{\citenamefont {Hellman}\ \emph {et~al.}(2011)\citenamefont
  {Hellman}, \citenamefont {Abrikosov},\ and\ \citenamefont
  {Simak}}]{Hellman2011}%
  \BibitemOpen
  \bibfield  {author} {\bibinfo {author} {\bibfnamefont {O.}~\bibnamefont
  {Hellman}}, \bibinfo {author} {\bibfnamefont {I.~A.}\ \bibnamefont
  {Abrikosov}}, \ and\ \bibinfo {author} {\bibfnamefont {S.~I.}\ \bibnamefont
  {Simak}},\ }\href {\doibase 10.1103/PhysRevB.84.180301} {\bibfield  {journal}
  {\bibinfo  {journal} {Physical Review B}\ }\textbf {\bibinfo {volume} {84}},\
  \bibinfo {pages} {180301} (\bibinfo {year} {2011})}\BibitemShut {NoStop}%
\bibitem [{\citenamefont {Hellman}\ \emph {et~al.}(2013)\citenamefont
  {Hellman}, \citenamefont {Steneteg}, \citenamefont {Abrikosov},\ and\
  \citenamefont {Simak}}]{Hellman2013}%
  \BibitemOpen
  \bibfield  {author} {\bibinfo {author} {\bibfnamefont {O.}~\bibnamefont
  {Hellman}}, \bibinfo {author} {\bibfnamefont {P.}~\bibnamefont {Steneteg}},
  \bibinfo {author} {\bibfnamefont {I.~a.}\ \bibnamefont {Abrikosov}}, \ and\
  \bibinfo {author} {\bibfnamefont {S.~I.}\ \bibnamefont {Simak}},\ }\href
  {\doibase 10.1103/PhysRevB.87.104111} {\bibfield  {journal} {\bibinfo
  {journal} {Physical Review B}\ }\textbf {\bibinfo {volume} {87}},\ \bibinfo
  {pages} {104111} (\bibinfo {year} {2013})}\BibitemShut {NoStop}%
\bibitem [{\citenamefont {Omini}\ and\ \citenamefont
  {Sparavigna}(1997)}]{Omini}%
  \BibitemOpen
  \bibfield  {author} {\bibinfo {author} {\bibfnamefont {M.}~\bibnamefont
  {Omini}}\ and\ \bibinfo {author} {\bibfnamefont {A.}~\bibnamefont
  {Sparavigna}},\ }\href@noop {} {\bibfield  {journal} {\bibinfo  {journal}
  {Nuovo Cimento della Societa Italiana di Fisica D}\ }\textbf {\bibinfo
  {volume} {19D}},\ \bibinfo {pages} {1537} (\bibinfo {year}
  {1997})}\BibitemShut {NoStop}%
\bibitem [{\citenamefont {Pascual-Gutierrez}\ \emph {et~al.}(2009)\citenamefont
  {Pascual-Gutierrez}, \citenamefont {Murthy},\ and\ \citenamefont
  {Viskanta}}]{Pascual-Gutierrez2009}%
  \BibitemOpen
  \bibfield  {author} {\bibinfo {author} {\bibfnamefont {J.~a.}\ \bibnamefont
  {Pascual-Gutierrez}}, \bibinfo {author} {\bibfnamefont {J.~Y.}\ \bibnamefont
  {Murthy}}, \ and\ \bibinfo {author} {\bibfnamefont {R.}~\bibnamefont
  {Viskanta}},\ }\href {\doibase 10.1063/1.3195080} {\bibfield  {journal}
  {\bibinfo  {journal} {Journal of Applied Physics}\ }\textbf {\bibinfo
  {volume} {106}},\ \bibinfo {pages} {063532} (\bibinfo {year}
  {2009})}\BibitemShut {NoStop}%
\bibitem [{\citenamefont {Kresse}(1999)}]{Kresse1999}%
  \BibitemOpen
  \bibfield  {author} {\bibinfo {author} {\bibfnamefont {G.}~\bibnamefont
  {Kresse}},\ }\href {\doibase 10.1103/PhysRevB.59.1758} {\bibfield  {journal}
  {\bibinfo  {journal} {Physical Review B}\ }\textbf {\bibinfo {volume} {59}},\
  \bibinfo {pages} {1758} (\bibinfo {year} {1999})}\BibitemShut {NoStop}%
\bibitem [{\citenamefont {Kresse}\ and\ \citenamefont
  {Furthm\"{u}ller}(1996)}]{Kresse1996}%
  \BibitemOpen
  \bibfield  {author} {\bibinfo {author} {\bibfnamefont {G.}~\bibnamefont
  {Kresse}}\ and\ \bibinfo {author} {\bibfnamefont {J.}~\bibnamefont
  {Furthm\"{u}ller}},\ }\href {\doibase 10.1103/PhysRevB.54.11169} {\bibfield
  {journal} {\bibinfo  {journal} {Physical Review B}\ }\textbf {\bibinfo
  {volume} {54}},\ \bibinfo {pages} {11169} (\bibinfo {year}
  {1996})}\BibitemShut {NoStop}%
\bibitem [{\citenamefont {Kresse}\ and\ \citenamefont
  {Hafner}(1993)}]{Kresse1993b}%
  \BibitemOpen
  \bibfield  {author} {\bibinfo {author} {\bibfnamefont {G.}~\bibnamefont
  {Kresse}}\ and\ \bibinfo {author} {\bibfnamefont {J.}~\bibnamefont
  {Hafner}},\ }\href {\doibase 10.1103/PhysRevB.48.13115} {\bibfield  {journal}
  {\bibinfo  {journal} {Physical Review B}\ }\textbf {\bibinfo {volume} {48}},\
  \bibinfo {pages} {13115} (\bibinfo {year} {1993})}\BibitemShut {NoStop}%
\bibitem [{\citenamefont {Kresse}(1996)}]{Kresse1996c}%
  \BibitemOpen
  \bibfield  {author} {\bibinfo {author} {\bibfnamefont {G.}~\bibnamefont
  {Kresse}},\ }\href {\doibase 10.1016/0927-0256(96)00008-0} {\bibfield
  {journal} {\bibinfo  {journal} {Computational Materials Science}\ }\textbf
  {\bibinfo {volume} {6}},\ \bibinfo {pages} {15} (\bibinfo {year}
  {1996})}\BibitemShut {NoStop}%
\bibitem [{\citenamefont {Nos\'{e}}(1984)}]{Nose1984}%
  \BibitemOpen
  \bibfield  {author} {\bibinfo {author} {\bibfnamefont {S.}~\bibnamefont
  {Nos\'{e}}},\ }\href {\doibase 10.1080/00268978400101201} {\bibfield
  {journal} {\bibinfo  {journal} {Molecular Physics}\ }\textbf {\bibinfo
  {volume} {52}},\ \bibinfo {pages} {255} (\bibinfo {year} {1984})}\BibitemShut
  {NoStop}%
\bibitem [{\citenamefont {Perdew}\ \emph {et~al.}(1996)\citenamefont {Perdew},
  \citenamefont {Burke},\ and\ \citenamefont {Ernzerhof}}]{Perdew1996}%
  \BibitemOpen
  \bibfield  {author} {\bibinfo {author} {\bibfnamefont {J.~P.}\ \bibnamefont
  {Perdew}}, \bibinfo {author} {\bibfnamefont {K.}~\bibnamefont {Burke}}, \
  and\ \bibinfo {author} {\bibfnamefont {M.}~\bibnamefont {Ernzerhof}},\ }\href
  {\doibase 10.1103/PhysRevLett.77.3865} {\bibfield  {journal} {\bibinfo
  {journal} {Physical Review Letters}\ }\textbf {\bibinfo {volume} {77}},\
  \bibinfo {pages} {3865} (\bibinfo {year} {1996})}\BibitemShut {NoStop}%
\bibitem [{\citenamefont {Weinstein}\ and\ \citenamefont
  {Piermarini}(1975)}]{Weinstein1975}%
  \BibitemOpen
  \bibfield  {author} {\bibinfo {author} {\bibfnamefont {B.}~\bibnamefont
  {Weinstein}}\ and\ \bibinfo {author} {\bibfnamefont {G.}~\bibnamefont
  {Piermarini}},\ }\href {\doibase 10.1103/PhysRevB.12.1172} {\bibfield
  {journal} {\bibinfo  {journal} {Physical Review B}\ }\textbf {\bibinfo
  {volume} {12}},\ \bibinfo {pages} {1172} (\bibinfo {year}
  {1975})}\BibitemShut {NoStop}%
\bibitem [{\citenamefont {Fabian}\ and\ \citenamefont
  {Allen}(1997)}]{Fabian1997}%
  \BibitemOpen
  \bibfield  {author} {\bibinfo {author} {\bibfnamefont {J.}~\bibnamefont
  {Fabian}}\ and\ \bibinfo {author} {\bibfnamefont {P.}~\bibnamefont {Allen}},\
  }\href {\doibase 10.1103/PhysRevLett.79.1885} {\bibfield  {journal} {\bibinfo
   {journal} {Physical Review Letters}\ }\textbf {\bibinfo {volume} {79}},\
  \bibinfo {pages} {1885} (\bibinfo {year} {1997})}\BibitemShut {NoStop}%
\bibitem [{\citenamefont {Broido}\ \emph {et~al.}(2005)\citenamefont {Broido},
  \citenamefont {Ward},\ and\ \citenamefont {Mingo}}]{Broido2005}%
  \BibitemOpen
  \bibfield  {author} {\bibinfo {author} {\bibfnamefont {D.}~\bibnamefont
  {Broido}}, \bibinfo {author} {\bibfnamefont {a.}~\bibnamefont {Ward}}, \ and\
  \bibinfo {author} {\bibfnamefont {N.}~\bibnamefont {Mingo}},\ }\href
  {\doibase 10.1103/PhysRevB.72.014308} {\bibfield  {journal} {\bibinfo
  {journal} {Physical Review B}\ }\textbf {\bibinfo {volume} {72}},\ \bibinfo
  {pages} {1} (\bibinfo {year} {2005})}\BibitemShut {NoStop}%
\bibitem [{\citenamefont {Wallace}(1998)}]{wallace1998thermodynamics}%
  \BibitemOpen
  \bibfield  {author} {\bibinfo {author} {\bibfnamefont {D.~C.}\ \bibnamefont
  {Wallace}},\ }\href {http://books.google.se/books?id=qLzOmwSgMIsC} {\emph
  {\bibinfo {title} {{Thermodynamics of crystals}}}},\ Dover Books on Physics\
  (\bibinfo  {publisher} {Dover Publications, Incorporated},\ \bibinfo {year}
  {1998})\BibitemShut {NoStop}%
\bibitem [{\citenamefont {Monkhorst}\ and\ \citenamefont
  {Pack}(1976)}]{Monkhorst1976a}%
  \BibitemOpen
  \bibfield  {author} {\bibinfo {author} {\bibfnamefont {H.}~\bibnamefont
  {Monkhorst}}\ and\ \bibinfo {author} {\bibfnamefont {J.}~\bibnamefont
  {Pack}},\ }\href {\doibase 10.1103/PhysRevB.13.5188} {\bibfield  {journal}
  {\bibinfo  {journal} {Physical Review B}\ }\textbf {\bibinfo {volume} {13}},\
  \bibinfo {pages} {5188} (\bibinfo {year} {1976})}\BibitemShut {NoStop}%
\bibitem [{\citenamefont {Yates}\ \emph {et~al.}(2007)\citenamefont {Yates},
  \citenamefont {Wang}, \citenamefont {Vanderbilt},\ and\ \citenamefont
  {Souza}}]{Yates2007}%
  \BibitemOpen
  \bibfield  {author} {\bibinfo {author} {\bibfnamefont {J.}~\bibnamefont
  {Yates}}, \bibinfo {author} {\bibfnamefont {X.}~\bibnamefont {Wang}},
  \bibinfo {author} {\bibfnamefont {D.}~\bibnamefont {Vanderbilt}}, \ and\
  \bibinfo {author} {\bibfnamefont {I.}~\bibnamefont {Souza}},\ }\href
  {\doibase 10.1103/PhysRevB.75.195121} {\bibfield  {journal} {\bibinfo
  {journal} {Physical Review B}\ }\textbf {\bibinfo {volume} {75}},\ \bibinfo
  {pages} {195121} (\bibinfo {year} {2007})}\BibitemShut {NoStop}%
\bibitem [{\citenamefont {Esfarjani}\ \emph {et~al.}(2011)\citenamefont
  {Esfarjani}, \citenamefont {Chen},\ and\ \citenamefont
  {Stokes}}]{Esfarjani2011}%
  \BibitemOpen
  \bibfield  {author} {\bibinfo {author} {\bibfnamefont {K.}~\bibnamefont
  {Esfarjani}}, \bibinfo {author} {\bibfnamefont {G.}~\bibnamefont {Chen}}, \
  and\ \bibinfo {author} {\bibfnamefont {H.}~\bibnamefont {Stokes}},\ }\href
  {\doibase 10.1103/PhysRevB.84.085204} {\bibfield  {journal} {\bibinfo
  {journal} {Physical Review B}\ }\textbf {\bibinfo {volume} {84}} (\bibinfo
  {year} {2011}),\ 10.1103/PhysRevB.84.085204}\BibitemShut {NoStop}%
\bibitem [{\citenamefont {Delaire}\ \emph {et~al.}(2011)\citenamefont
  {Delaire}, \citenamefont {Marty}, \citenamefont {Stone}, \citenamefont
  {Kent}, \citenamefont {Lucas}, \citenamefont {Abernathy}, \citenamefont
  {Mandrus},\ and\ \citenamefont {Sales}}]{Delaire2011a}%
  \BibitemOpen
  \bibfield  {author} {\bibinfo {author} {\bibfnamefont {O.}~\bibnamefont
  {Delaire}}, \bibinfo {author} {\bibfnamefont {K.}~\bibnamefont {Marty}},
  \bibinfo {author} {\bibfnamefont {M.~B.}\ \bibnamefont {Stone}}, \bibinfo
  {author} {\bibfnamefont {P.~R.~C.}\ \bibnamefont {Kent}}, \bibinfo {author}
  {\bibfnamefont {M.~S.}\ \bibnamefont {Lucas}}, \bibinfo {author}
  {\bibfnamefont {D.~L.}\ \bibnamefont {Abernathy}}, \bibinfo {author}
  {\bibfnamefont {D.}~\bibnamefont {Mandrus}}, \ and\ \bibinfo {author}
  {\bibfnamefont {B.~C.}\ \bibnamefont {Sales}},\ }\href {\doibase
  10.1073/pnas.1014869108} {\bibfield  {journal} {\bibinfo  {journal}
  {Proceedings of the National Academy of Sciences}\ }\textbf {\bibinfo
  {volume} {108}},\ \bibinfo {pages} {4725} (\bibinfo {year}
  {2011})}\BibitemShut {NoStop}%
\bibitem [{\citenamefont {Nyhus}\ \emph {et~al.}(1995)\citenamefont {Nyhus},
  \citenamefont {Cooper},\ and\ \citenamefont {Fisk}}]{Nyhus1995}%
  \BibitemOpen
  \bibfield  {author} {\bibinfo {author} {\bibfnamefont {P.}~\bibnamefont
  {Nyhus}}, \bibinfo {author} {\bibfnamefont {S.}~\bibnamefont {Cooper}}, \
  and\ \bibinfo {author} {\bibfnamefont {Z.}~\bibnamefont {Fisk}},\ }\href
  {\doibase 10.1103/PhysRevB.51.15626} {\bibfield  {journal} {\bibinfo
  {journal} {Physical Review B}\ }\textbf {\bibinfo {volume} {51}},\ \bibinfo
  {pages} {15626} (\bibinfo {year} {1995})}\BibitemShut {NoStop}%
\end{thebibliography}

%

\end{document}